\documentclass[prl,aps,showpacs,epsf,twocolumn]{revtex4}

\usepackage{graphicx}
\usepackage{amsmath}
\usepackage{amssymb}
\usepackage{amsthm}

\begin{document}

\title{Trap and population imbalanced two-component Fermi gas in the BEC limit}

\author{S. A. Silotri}

\email{silotri@prl.res.in}

\affiliation{Theoretical Physics Division, Physical Research Laboratory, Ahmedabad 380009, India}

\date{\today}

\begin{abstract}
We study equal mass population imbalanced two-component atomic Fermi gas with unequal trap 
frequencies $(\omega_{\uparrow} \neq \omega_{\downarrow})$ at zero temperature 
using the local density approximation (LDA). We consider the strongly attracting Bose-Einstein condensation 
(BEC) limit where polarized (gapless) superfluid is stable. The system exhibits shell 
structure: unpolarized SF$\rightarrow$polarized SF$\rightarrow$normal N.
Compared to trap symmetric case, when the majority component is tightly confined
the gapless superfluid shell grows in size
leading to reduced threshold polarization to form polarized (gapless) superfluid core.
In contrast, when the minority component is tightly confined, we find that
the superfluid phase is dominated by unpolarized superfluid phase with gapless phase 
forming a narrow shell.
The shell radii for various phases as a function of polarization at different values 
of trap asymmetry are presented and the features are explained using 
the phase diagram.

\end{abstract}


\maketitle

\section{Introduction}

Ultracold atomic Fermi gases present a unique opportunity to study the exotic pairing phases
where the effective interaction is tunable via Feshbach resonance and 
population of each spin component can be controlled.
These studies with two-component Fermi gas include the 
Bardeen-Cooper-Schrieffer (BCS) to Bose-Einstein condensation (BEC) 
crossover~\cite{stringari_RMP, kett_jpc} with equal population two-component Fermi gas and 
the effects of population imbalance on the superfluid state~\cite{kett_jpc,kett_1,hulet,kett_2,kett_prl,kett_nopair,kett_resonant, kett_BF}.
For Fermi gas with population imbalance, various pairing 
scenarios are proposed: Fulde-Ferrel-Larkin-Ovchinnikov phase FFLO~\cite{FF,LO}, 
breached pairing~\cite{wilz}, phase separation~\cite{caldas} and 
pairing with deformed fermi surface~\cite{sedrakian}.

The present studies of equal mass superfluid Fermi systems involve same trapping potentials for
the two component.
The  zero~\cite{sheehy_prl} and finite temperature~\cite{parish_natphys} phase diagrams of the equal 
mass population imbalanced system taking into account various pairing phases with 
implications to experiments are well understood.
The trap imbalanced  is naturally realized with Fermi mixture with unequal 
masses where each component experiences different potential due to the mass difference. 
The ground state properties for this system have been studied in 
Ref.~\cite{duan_2006, iskin_mass, parish_mass, yip_mass}. However, it was recently 
proposed in Ref.~\cite{iskin} that even equal mass Fermi mixture 
can admit trap imbalance and the system was studied with population balance~\cite{iskin}
and small trap imbalance~\cite{blume}.

Motivated by recent experiment performed in the BEC regime of interaction exploring Bose-Fermi mixture~\cite{kett_BF}, we consider the trap and population imbalanced  $(\omega_{\uparrow} \neq \omega_{\downarrow})$ 
Fermi mixture in this regime. The phase diagram  in this regime becomes richer with the 
existence of the gapless (polarized) superfluid also referred to as 
breached pair phase with one fermi surface. We study the effects of trap asymmetry on the shell structure as
function of polarization.
The shell structure, in general, consists of three phases: unpolarized superfluid (BCS SF) at the center, 
the breached pair phase with one fermi surface (BP1) forms the intermediate shell finally
surrounded by polarized normal (N). We do not consider the FFLO  phase~\cite{FF, LO}
where cooper pair carries finite center-of-mass momentum. 
This phase is stable within very narrow window of the 
applied chemical potential difference in the BCS regime.


\section{Formalism}
The system we consider is a trapped cloud  of two-component Fermionic mixture confined  
by harmonic isotropic potential $V_{T\sigma}(r)$ where $r$ measures the distance 
from the trap center. The Fermi atoms have unequal population of the 
two pseudo-spin (hyperfine) states and interact via point-contact $s$-wave interaction.
To proceed further we start with system without the trap and later include it under 
local density approximation. 
The Hamiltonian density ($\hbar=1$) for the fermions in this case is given by
\begin{equation}
  H=\sum_{\sigma}\Psi_{\sigma}^{^{\dagger}}(\mathbf{r})\left(\varepsilon_{k\sigma}-
  \mu_{\sigma}\right)\Psi_{\sigma}(\mathbf{r})+g\Psi_{\uparrow}^{^{\dagger}}(\mathbf{r})
  \Psi_{\downarrow}^{^{\dagger}}(\mathbf{r})\Psi_{\downarrow}(\mathbf{r})\Psi_{\uparrow}(\mathbf{r}),
 \label{hamiltonian}
\end{equation}
where $\Psi_{\sigma}^{^{\dagger}}(\mathbf{r})$ creates a pseudospin-$\sigma$ 
fermion at position $\mathbf{r}, $ $\varepsilon_{k\sigma}=k^{2}/2m_{\sigma};$ 
$\mu_{\sigma}$ and $m_{\sigma}$ are the  chemical potential and mass for 
pseudospin-$\sigma$ component with $\sigma=\uparrow,\downarrow$.
$g$ is the bare coupling constant characterizing interparticle interaction.
It is related to $s$-wave scattering length $a$ of the system by the Lippmann-Schwinger 
equation
\begin{equation}
 \frac{\tilde{m}}{2\pi a}= \frac{1}{g}+
  \frac{1}{V}\sum_{\mathbf{k}}\frac{1}{2\varepsilon_{\mathbf{k}}},
  \label{lippmann}
\end{equation}
where the average kinetic energy $\varepsilon_{\mathbf{k}}=(\varepsilon_{1\mathbf{k}}+\varepsilon_{2\mathbf{k}})/2,$  $V$ is volume and reduced mass $\tilde m=m{\uparrow}m_{\downarrow}/(m{\uparrow}+m_{\downarrow}).$
The  pairing Hamiltonian can be diagonalized  by  Bogoliubov transformations 
using thermofield dynamics techniques as described in \cite{mypaper, mishra}. 
This formalism also accounts for the  polarized
superfluidity  when two components have unequal population or masses.
This leads to  thermodynamic potential density,

\begin{equation}
  \Omega=\frac{1}{V}\sum_{\mathbf{k}}\left[\xi_{\mathbf{k}}-E_{\mathbf{k}}-\frac{1}{\beta}
         \sum_{\sigma}\ln\left(1+\exp\left(-\beta E_{\sigma}\right)
         \right)\right]-\frac{\Delta^{2}}{g},
	 \label{thermo_potential}
\end{equation}
where we have defined  gap or order parameter $\Delta(\mathbf{r})=-g\left\langle\Psi_{\downarrow}(\mathbf{r})\Psi_{\uparrow}(\mathbf{r})\right\rangle.$
Introducing the chemical potential $\mu=\left(\mu_{\uparrow}+\mu_{\downarrow}\right)/2$ 
and the chemical potential difference or the Zeeman field $h=\left(\mu_{\uparrow}-\mu_{\downarrow}\right)/2$, 
we define  quasiparticle energies $E_{\sigma}=E\pm\delta\xi$ with
$\xi_{\mathbf{k}}=\varepsilon_{\mathbf{k}}-\mu,$
$E_{\mathbf{k}}=\sqrt{\xi_{\mathbf{k}}^{2}+\Delta^{2}},$
for equal mass case $\delta\xi=-h$
and $\beta=1/T.$ 

Note that the there are two branches for the quasiparticle energies and lead to 
gapless modes when $E_{k\uparrow}=0$ assuming $\uparrow$-fermions to be majority.
At zero temperature the excess fermions are accommodated in negative quasiparticle energy states.
This gapless phase is referred to as breached pair phase with one Fermi surface (BP1) 
as it is stable only for $\mu <0$ and hence possesses one Fermi surface.
In the deep BEC regime this phase can be understood as mixture of composite bosons and 
fermion quasiparticles~\cite{parish_natphys}.

The gap equation is given by the condition of extremum of thermodynamic
potential density $\frac{\partial\Omega}{\partial\Delta}=0$. The
average number density and density difference equation respectively are given by $n= -\frac{\partial\Omega}{\partial\mu}$ and  $m=-\frac{\partial\Omega}{\partial h}.$

Before considering the trapped system, we construct the zero temperature phase diagram
in grand canonical ensemble~\cite{sheehy_review} of fixed $\mu$ and $h.$

\begin{figure}
\includegraphics[bb= 57 47 686 513,width=\columnwidth, clip=]{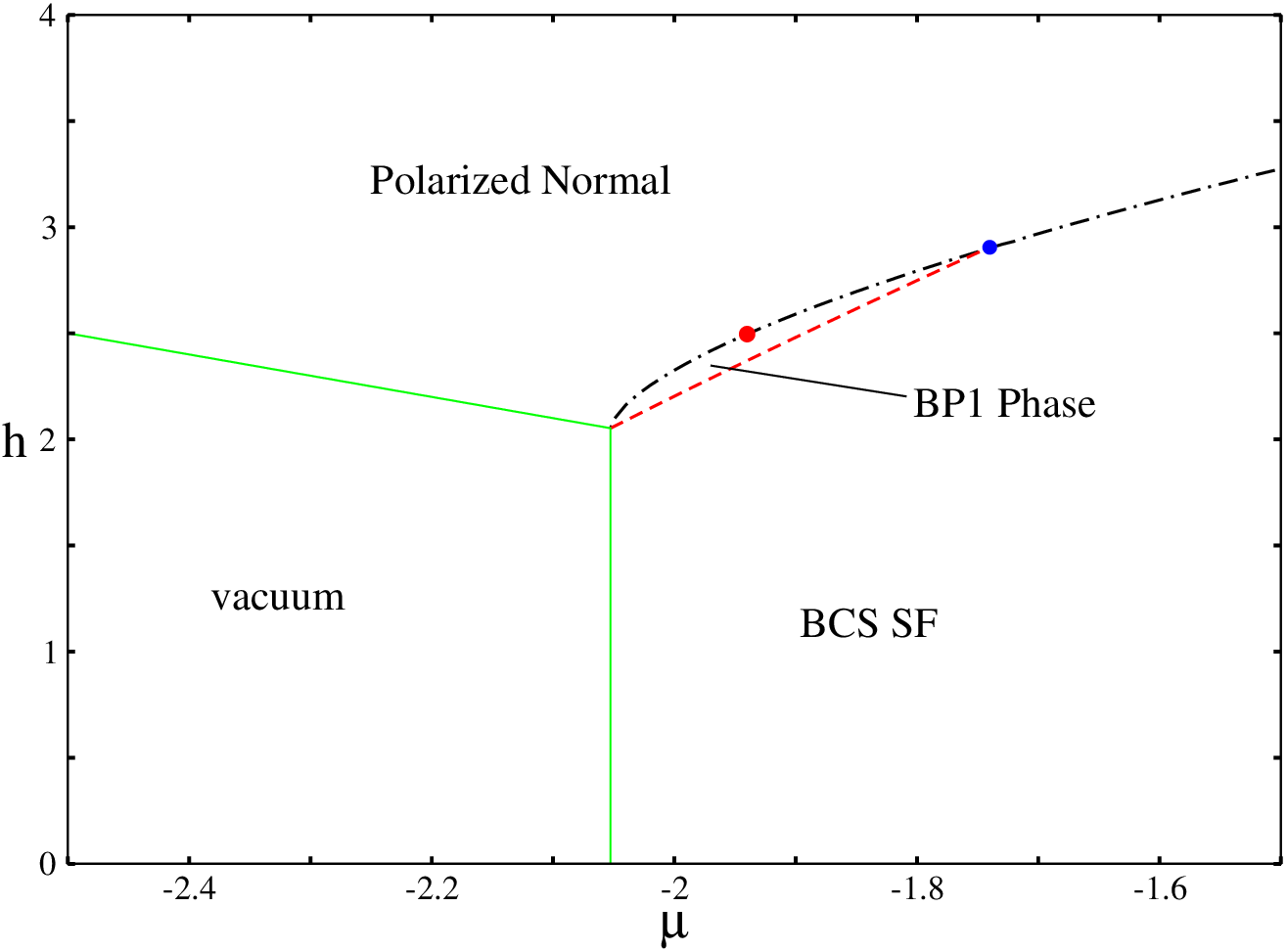}
\caption{(Color online) The zero temperature phase diagram for $(k_Fa)^-1=2.0$
showing unpolarized superfluid (BCS SF), polarized superfluid (BP1), vaccum 
and polarized normal (N) phases. The upper (blue) dot denotes the point 
beyond which BP1 state ceases to exist. 
The lower (red) dot represents the tricritical point. 
The dashed(red) line indicates the second-order transition between unpolarized Sf and  BP1 phase.
The dot dashed (black) line indicates first-order transition between SF to polarized N state
and BP1 to normal above and  below the upper (blue) point respectively. }
\label{lda_phase}
\end{figure}
%
We start with $h=0$ and find the point where superfluid state make continuous transition to
vacuum state of molecules. This value of $\mu$ is denoted by $\mu_c.$
For small $h<h_m$ this behavior i.e. superfluid-to-Vacuum persists and leads to vertical
phase boundary in the diagram.

Next, we start with $\mu<\mu_c$ and increasing $h.$ 
Here system evolve from vacuum state to polarized normal state as $\mu_{\uparrow}=\mu+h$ 
is now positive quantity leading to finite population of the $\uparrow$-fermions.
There cannot be superfluid phase here as $\mu<\mu_c.$
This leads to Vaccum-to-Polarized N phase boundary.

Similarly we start with $\mu>\mu_c$ with increasing $h.$
Here system makes continuous transition to Breached pair state as 
superfluid starts to admit finite polarization.
The superfluid-to BP1 boundray is calculated by numerical comparison 
of the thermodynamic potential in the respective states.
As we further increase the $h$, the BP1 eventually make transition 
to polarized normal state. However, depending on the value of $\mu$ the BP1-to-Polarized 
transition can be first or second order.
The tricritical point where first and second order transition meet is indicated in the phase diagram.
Also the BP1-Normal first order curve intersect the superfluid-BP1 curve at large $\mu.$ Beyond 
this intersection point, BP1 ceases to exist and there is direct first order superfluid-to-Normal transition.

\section{Trapped Fermi mixture}\label{trapped}
We next consider the trapped fermions confined by harmonic isotropic potential.
This can be taken into account via local density approximation (LDA) where trapped system is treated
as locally uniform with local chemical potential $\mu_{\sigma}(r)=\mu_{\sigma}-V_{T\sigma}.$
$\mu_{\sigma}$ is actual Lagrange multiplier constraining number of atoms and $V_{T\sigma}=\frac{1}{2}m_{\sigma}\Omega_{\sigma}r^2$ with
$\omega_{\sigma}$ as trapping frequency for component $\sigma.$
Furthermore, the chemical potentials of each component at a given point in the trap 
can be written as 
\begin{align}
 \mu_1(r)&=\mu(r)+h(r)\\
 \mu_2(r)&=\mu(r)-h(r)
\end{align}
where $\mu(r)=\mu-V_T(r)$ and $h(r)=h-\delta V_T(r)$ where $V_T(r)=(V_{T\uparrow}+V_{T\downarrow})/2$
and $\delta V_T(r)=(V_{T\uparrow}-V_{T\downarrow})/2.$
The quantities $\mu$ and $h$ are determined by enforcing particle number 
constraints namely total number of atoms and population difference respectively.

\begin{figure}
\includegraphics[bb=61 61 385 513,width=\columnwidth, clip=,]{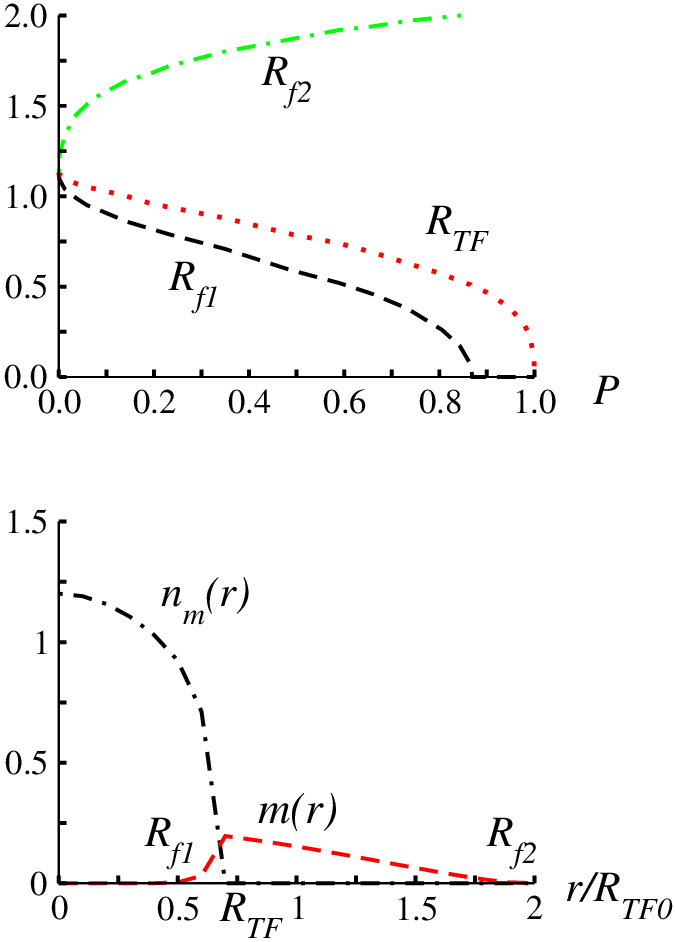}
\caption{(Color online) (a) The three radii $R_{f1}$ (outer boundary of unpolarized superfluid),
$R_{TF}$ (outer boundary of BP-1 phase) and $R_{f2}$ (outer boundary of N phase) plotted
as a function of polarization $P$ at trap asymmetry parameter $\eta=0$ and $(k_Fa)^{-1}=2.0.$
(b) The molecular density $n_m$ and magnetization $m$ plotted against radius $r$ measured in units of 
$k_F^3$ and $R_{TF0}.$}
\label{fig:0all}
\end{figure}

In terms of $n(r)=n_{\uparrow}(r)+n_{\downarrow}(r)$ and $m(r)=n_{\uparrow}(r)-n_{\downarrow}(r)$, 
the total number of atoms $N$ and population imbalance $\Delta N$ are given by
$N=\int d^3r\,n(r)$ and $\Delta N=\int d^3r\,m(r)$ where
\begin{eqnarray}
   n_{\uparrow}(r) & = & \frac{1}{(2\pi)^3}\int d^3k\left[\frac{1}{2}
                \left(1+\frac{\xi(r)}{E(r)}\right)\Theta(-E_{\uparrow}(r)) \right.
                      \nonumber \\
           &+&   \left. \frac{1}{2}\left(1-\frac{\xi(r)}{E(r)}\right)(1-\Theta(-
                E_{\downarrow}(r)))\right]
                 \\
   n_{\downarrow}(r) & = & \frac{1}{(2\pi)^3}\int d^3k\left[\frac{1}{2}
                \left(1+\frac{\xi(r)}{E(r)}\right)\Theta(-E_{\downarrow}(r)) \right.
                      \nonumber \\
           &+&   \left. \frac{1}{2}\left(1-\frac{\xi(r)}{E(r)}\right)(1-\Theta(-
                E_{\uparrow}(r)))\right]. \label{eq:n12}
\end{eqnarray}
where $\Theta(\ldots)$ is the Heaviside step function, the zero-temperature limit for Fermi-Dirac distribution.
The local gap equation is 
\begin{eqnarray}
   -\frac{\tilde{m}}{2\pi a}&=&\frac{1}{(2\pi)^3}
          \int d^3k\left [ \frac{1}{2E(r)}\big(1-\Theta(-E_{\uparrow}(r))\right.
           \nonumber\\
         && \left. - \Theta(-E_{\downarrow}(r)) \big) -\frac{1}{2\varepsilon_{k}}\right].
  \label{eq:gap}
\end{eqnarray}

The trap introduces the new length scale
called Thomas-Fermi radius defined as $R_{TF}=\sqrt{\frac{2\mu}{m\omega_{T}^{2}}}.$
Note further that in the BEC regime,  chemical potential $\mu$ is
already negative at the center of the trap and hence $\mu(r)$ does not vanish. 
We also note that in the deep BEC regime $\mu=-E_b/2$~\cite{randeria}, the molecular binding energy
with $E_b=1/2m_ra^2$. Thus we impose the condition~\cite{sheehy_review},
\begin{equation}
  \mu(R_TF0)=\mu_0-\frac{1}{2}m\omega_{T}^{2}R_{TF0}^{2}=-\frac{E_b}{2}.
\end{equation}
This gives,
\begin{equation}
 R_{TF0}=\sqrt{\frac{E_b(2\mu_0+1)}{m\Omega_T^2}}.
\end{equation}

The zero subscript indicates that the quantities are for zero polarization.
To investigate the system numerically we define the dimensionless quantities
$\hat\Delta(r)=\Delta/E_F,$ $\hat\mu(r)=\mu(r)/E_F,$ $\hat h(r)=h(r)/E_F$ where we choose
$E_F=(6N)^{1/3}\hbar\Omega_T$ with $\omega_T=\sqrt{\omega_{\uparrow}^2+\omega_{\downarrow}^2}$
We, also, normalize the distance in the trap $x=r/R_{TF0}$ and define $k_F$
by the relation $E_F=k_F^2/2\tilde m.$ Hence
\begin{equation}
\hat\mu_0(x)=\hat\mu_0 -x^2\left(\hat\mu_0+\frac{1}{(k_Fa)^2}\right),
\end{equation}
where we have expressed the binding energy $E_b$ in $E_F$ units as $E_b=E_F/(k_Fa)^2$ 
However for the system with population imbalance we have

\begin{align}
  \hat\mu(x)&=\hat\mu -\left(\hat\mu_0+\frac{1}{(k_Fa)^2}\right)x^2, \nonumber\\
  \hat h(x)&=\hat h -\eta\left(\hat\mu_0+\frac{1}{(k_Fa)^2}\right)x^2,
  \label{eq:hmu}
\end{align}
where dimensionless quantity $\eta=(\omega_{\uparrow}^2-\omega_{\downarrow}^2)/(\omega_{\uparrow}^2+\omega_{\downarrow}^2)$ controls the trap
asymmetry of the Fermi gas.

Now the equation for total number of atoms and population difference in dimensionless form
are given by
\begin{align}
  N &=\int d^3x\, n(\hat\mu(x),\hat h(x)),\\
 \Delta N&=\int d^3x\, m(\hat\mu(x),\hat h(x)).
\end{align}

The system for a given coupling strength and polarization $P=\Delta N/N$ is investigated numerically in the following manner: first $\mu_0$ is calculated by setting $P=0.$ Then using Eq.~\ref{eq:hmu} together with number
and population imbalance equation, $\mu$ and $h$ are calculated. In the BEC limit the order parameter $(\Delta)$
and density for composite bosons or molecular density are related~\cite{randeria}. By calculating the local composite boson density $(n_m)$ and magnetization $(m)$, the various phases are identified.

\begin{figure}
\includegraphics[bb=49 28 685 512,width=\columnwidth,clip=]{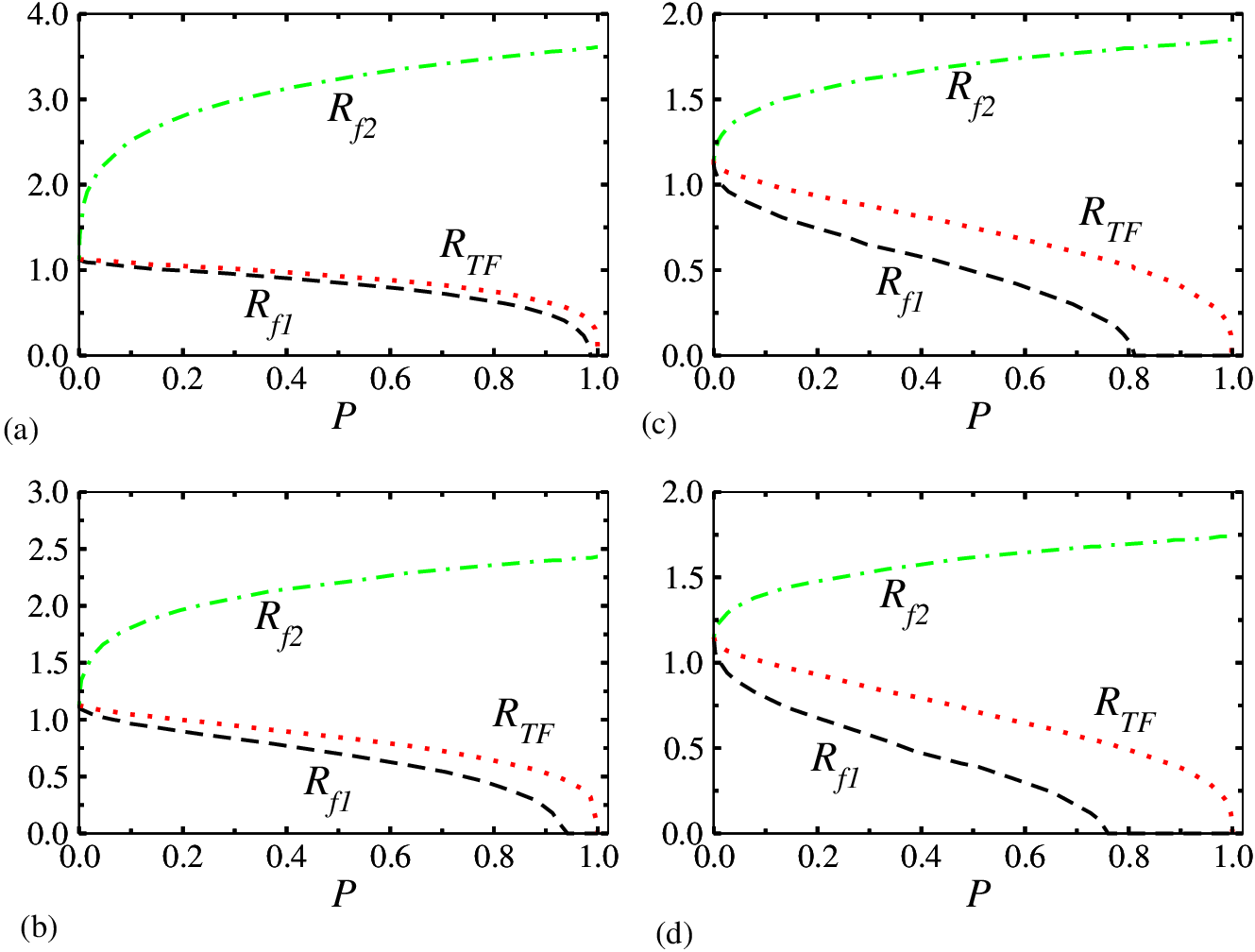}
\caption{(Color online) The three radii $R_{f1}$ (outer boundary of unpolarized superfluid),
$R_{TF}$ (outer boundary of BP1 phase) and $R_{f2}$ (outer boundary of N phase) plotted
as a function of polarization $P$ for various values of the trap asymmetry parameter
 $\eta.$ (a) $\eta=-0.9$, (b) $\eta=-0.5$
(c) $\eta=0.5$, (d) $\eta=0.9.$ All the radii are measured in units of $R_{TF0}$ 
(outer boundary of superfluid unpolarized  cloud).
}
\label{fig:radii}
\end{figure}

\section{Results and Discussions}

We choose experimentally accessible $(k_Fa)^{-1}=2.0$ for the interaction strength.
The phase at each spatial point of the trap is determined by the local chemical potentials
$\mu(r)$ and $h(r)$ (see Eqs.~\ref{eq:hmu}) mapping it 
to the corresponding point in the phase diagram.
As radius is increased $(\mu(r), h(r))$  moves towards left in the phase diagram 
forming a line segment.

We find three different phases in the cloud. At the center, superfluid core where the population 
of the two components are equal i.e. unpolarized superfluid (BCS SF), then an intermediate polarized 
superfluid (BP1) shell where fermion quasiparticles and and composite bosons coexist and finally
outer rim of majority component. This leads to three radii characterizing the shell structure:
$R_{f1}$ where $n_m \neq 0$ and $m(r)=0$ forming boundary for BCS SF phase, $R_{TF}$ above which
$n_m=0$ and $R_{f2}$ above which $m(r)=0.$ 

The three radii for the system without trap asymmetry $\eta=0$ as a function of polarization $P$
together with density profiles showing $n_m(r)$ and $m(r)$ are shown in Fig.~\ref{fig:0all}.
The shell structure consist of BCS SF phase for $r<R_{f1}$ , BP1 phase for $R_{f1}<r<R_{TF}.$ 
and finally polarized normal (N) state for $R_{TF}<r<R_{f2}.$

We next consider the system with trap asymmetry characterized by the dimensionless parameter
$\eta=(\omega_{\uparrow}^2-\omega_{\downarrow}^2)/(\omega_{\uparrow}^2+\omega_{\downarrow}^2).$
The positive (negative) $\eta$ value indicates that majority (minority) component 
is more tightly confined harmonically than the minority (majority) component. 

\begin{figure}
\includegraphics[bb=61 43 693 513,width=\columnwidth, clip=]{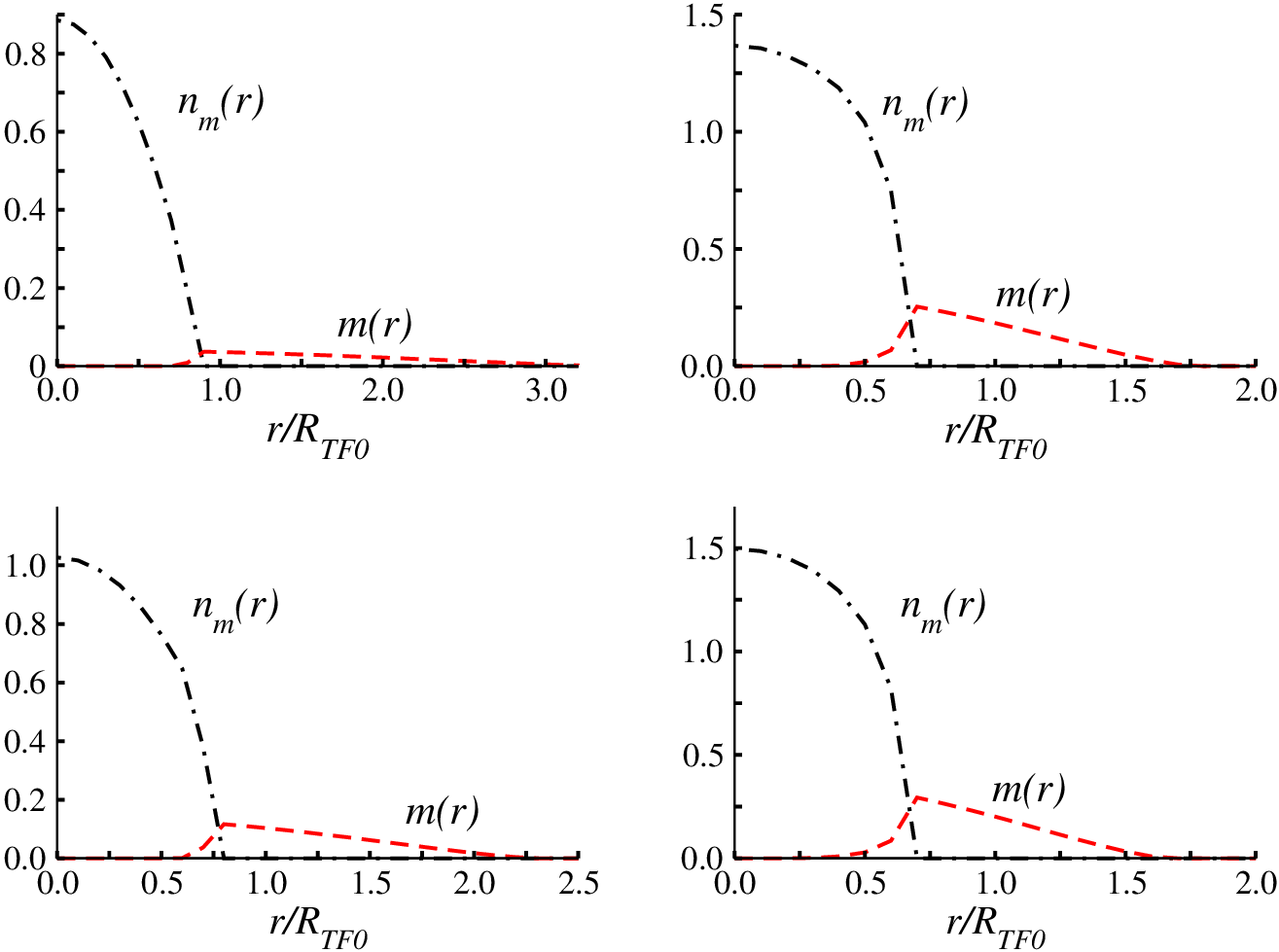}
\caption{(Color online) Density profiles at $P=0.65$ for different values 
of the trap asymmetry parameter $\eta$. (a) $\eta=-0.9$, (b) $\eta=-0.5$
(c) $\eta=0.5$, (d) $\eta=0.9.$ The molecular density $n_m$ and the 
magnetization plotted as a function of radius
measured in units of $k_F^3$ and $R_{TF0}$ respectively.}
\label{all_density}
\end{figure}

The three radii with different trap asymmetry parameter $\eta$ as functions of polarization
$P$ are shown in Fig.~\ref{fig:radii}. The value $\eta \pm 0.9$ corresponds to the situation 
when one of the component is very strongly confined.
We start with $\eta=-0.9$ corresponding to $\omega_{\uparrow} \ll\omega_{\downarrow}.$ 
The BP1 shell here is very narrow and overall size of the cloud ( characterized by $R_{f2}$) is much larger
than the superfluid cloud without population imbalance (the cloud size  is measured in units of $R_{TF0}$).
As we increase $\eta$, the BP1 shell grows in size, however, size of the cloud decreases.
The window of polarization for which BP1 phase forms the superfluid core starting at 
the center of the trap increases becoming maximum at $\eta=0.9$. The BP1 phase forms the core
beginning at $P=0.76,$ which should be experimentally feasible.
We also present the density profiles for same set of $\eta$ at $P=0.65$  in Fig.~\ref{all_density}.
We note that as $\eta$ increases, size of $n_m$ representing density of the composite bosons
shrinks but becomes more dense. It also exhibits the large cloud sizes for tightly confined minority
as noted above.

\begin{figure}
\includegraphics[bb=29 37 718 527,width=\columnwidth, clip=]{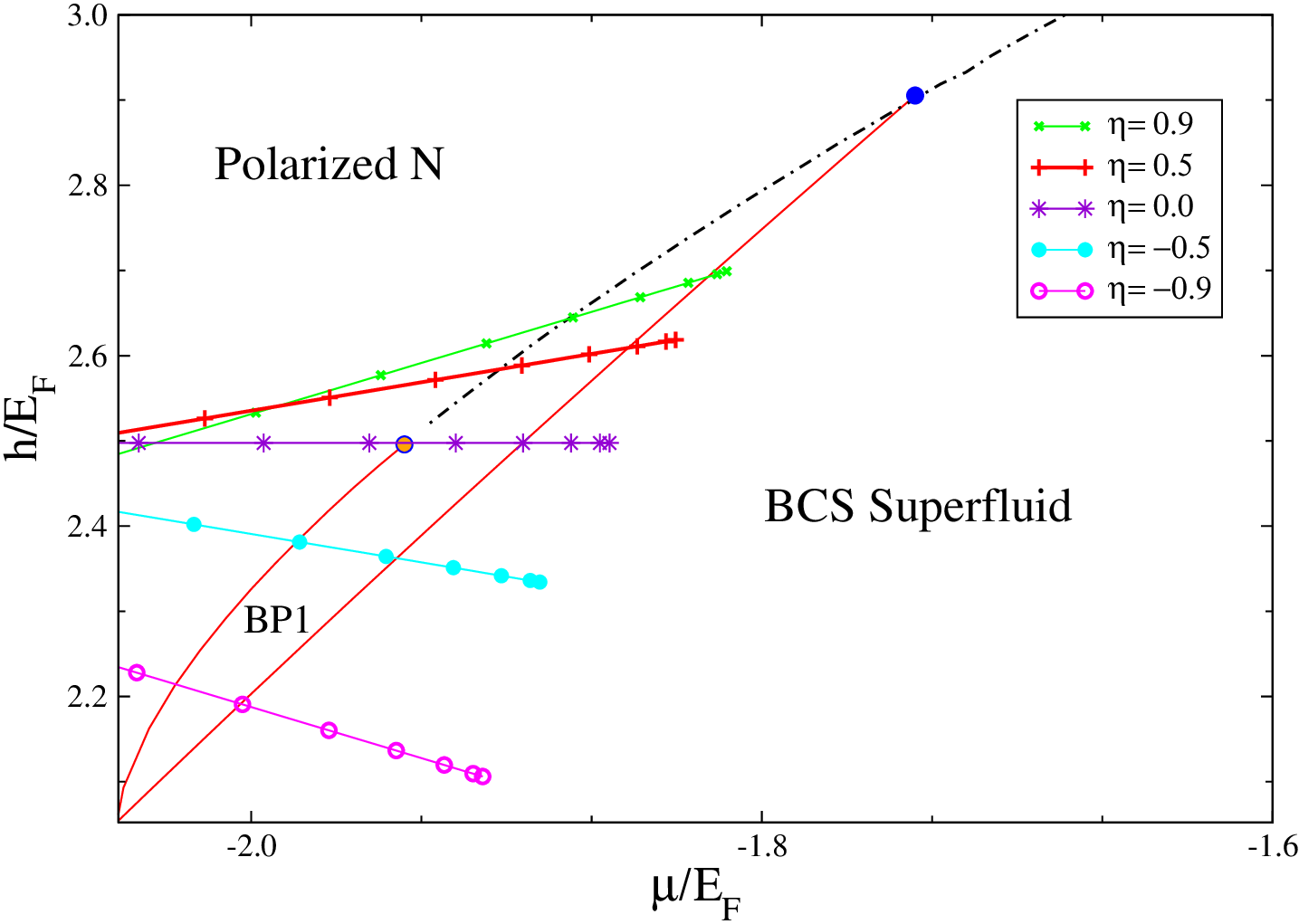}
\caption{(Color online) The part of phase diagram with the BP-1 region enlarged. 
The solid (red) line indicates the second order phase transition and 
dot dashed (black) line  shows first order transition. Superimposed are the $\mu-h$ 
variations as line segments for various values of the trap asymmetry parameter $\eta$ at 
$P=0.65$
}
\label{inset}
\end{figure}

All these features can be explained by analyzing ($\mu(r),h(r)$) variations for each value of $\eta$
in the phase diagram. To this end, we replot the phase diagram enlarging the BP1 state. 
The line segments representing above mentioned variations are also shown (Fig.~\ref{inset}).
We note that $\eta$ with positive (negative) value has a positive (negative) 
slopes with zero value for $\eta=0.$

Note further that BP1 to polarized N transition is second and first order for positive and 
negative set of chosen values respectively. These transitions can be detected via density profiles
in the experiments.

The $\eta=-0.9$ line segment traverses small region of BP1 phase making second order 
transition to polarized N while $\eta=0.9$ segment traverses larger region in 
BP-1 phase before making first order transition to polarized N.
The order of transition can be detected in experiments via spatial discontinuities
which vanish for second order transition. As $\eta$ 
is increased from $\eta=-0.9$ line segment increasingly  have larger portion 
in BP-1 region. This explains why the BP1 shell expands in size as $\eta$ is increased.
Note futher that owing to their negative (positive) slopes, $\eta$ with negative 
values have longer (shorter) excursion into the polarized N state before 
encountering the vacuum explaining the larger (smaller) size for their clouds.
Since all the atoms are integrated across the trap to conserve the atoms, the corresponding
atom density distributions are also affected accounting for increased (reduced) number of atoms in BP1 phase for
$\omega_{\uparrow} > \omega_{\downarrow}$ ($\omega_{\uparrow} < \omega_{\downarrow}$).

\section{Conclusion}
\label{Conclusions}
In conclusion, we have studied in this paper the asymmetrically trapped and 
population imbalanced two-components Fermi gas in the strongly attracting BEC limit at $T=0$. 
Using the local density approximation (LDA), we calculated the the shell radii of various 
phases in the trap as a function of polarization and trap asymmetry.
Compared to symmetric trap case $(\eta=0)$, we find that when the majority component is tightly confined the 
gapless superfluid shell (BP1) increases in size.
The polarization threshold to form the polarized BP1 superfluid at the core is reduced for a given interaction strength in this case. However, when minority are tightly confined 
unpolarized superfluid is favored with BP1 phase forming a narrow shell.
We explained these features using the phase diagram.

\acknowledgments
We wish to thank H. Mishra and D. Angom for discussions at the initial stages of this work.


\begin{references}

\bibitem{stringari_RMP} See, {\it e.g.}, S. Giorgini, L.P. Pitaevskii and S. Stringari, 
Rev. Mod. Phys. {\bf 80}, 1215 (2008) and references therein.

\bibitem{kett_jpc} W. Ketterle, Y. Shin, A. Schirotzek and C. H. Shunk 
J. Phys.: condens. Matter {\bf 21}, 164206 (2009).

\bibitem{kett_1} M.W. Zwierlein, A. Schirotzek, C.H. Schunck and W. Ketterle, Science {\bf 311}, 492 (2006).

\bibitem{hulet} G.B. Partridge, W. Li, R.I. Kamar, Y.A. Liao and R.G. Hulet, Science {\bf 311}, 503 (2006).


\bibitem{kett_2}M.W. Zwierlein, C.H. Schunck, A. Schirotzek, W. Ketterle,
{\bf Nature 442}, 54 (2006).



\bibitem{kett_prl} Y. Shin, M.W. Zwierlein, C.H. Schunck, A. Schirotzek and W. Ketterle, Phys. Rev. Lett. {\bf 97}, 030401 (2006).

\bibitem{kett_nopair} C. H. Schunck, Y. Shin, A. Schirotzek, M. W. Zwierlein,and W. Ketterle,
Science {\bf 316}, 867 (2007).

\bibitem{kett_resonant} Y. Shin, C.H. Schunck, A. Schirotzek and W. Ketterle, Nature {\bf 451}, 689 (2008).

\bibitem{kett_BF}Y. Shin, A. Schirotzek, C.H. Schunck, and W. Ketterle
Phys. Rev. Lett. {\bf 101}, 070404 (2008). 



\bibitem{FF} P. Fulde and R. A. Ferrell, Phys. Rev. \textbf{135}, A550 (1964).


\bibitem{LO} A. I. Larkin and Y. N. Ovchinnikov, Sov. Phys. JETP \textbf{20}, 762 (1965). 


\bibitem{wilz} W. V. Liu and F. Wilczek, Phys. Rev. Lett. {\bf 90}, 047002 (2003).

\bibitem{caldas} P. F. Bedaque, H. Caldas, and G. Rupak, Phys. Rev. Lett. 91, 247002 (2003).

\bibitem{sedrakian} A. Sedrakian, J. Mur-Petit, A. Polls, and H. Muther, Phys. Rev. A {\bf 72}, 013613 (2005).

\bibitem{sheehy_prl}D. E. Sheehy and L. Radzihovsky, Phys. Rev. Lett. {\bf 96}, 060401 (2006).

\bibitem{parish_natphys} M. M. Parish F. M. Marchetti, A. Lamacraft and B. D. Simons, 
Nature Phys. {\bf 3}, 124 (2007).


\bibitem{duan_2006} G. D Lin, W. Yi and L. M. Duan, Phys. Rev. A {\bf 74}, 031604(R) (2006).

\bibitem{iskin_mass} M. Iskin C. A. R. S\'{a} de Melo, Phys. Rev. Lett. {\bf 97}, 100404 (2006).

\bibitem{parish_mass} M. M. Parish, F. M. Marchetti,  A. Lamacraft and B. D. Simons,
Phys. Rev. Lett. {\bf 98}, 160402 (2007).

\bibitem{yip_mass} C. H. Pao, S. T. Wu and  S. K. Yip,
Phys. Rev. A {\bf 76}, 053621 (2007).


\bibitem{mypaper} S. Silotri, D. Angom, H. Mishra and A. Mishra, Eur. Phys. Jr. D {\bf 49}, 383-390 (2008).

\bibitem{mishra} H. Mishra and A. Mishra, Eur. Phys. Jr. D {\bf 53}, 75-87, (2009).

\bibitem{randeria}
C. A. R. S\'{a} de Melo, M. Randeria, and J. R. Engelbrecht, 
Phys. Rev. Lett. {\bf 71}, 3202 (1993).

\bibitem{sheehy_review} D. E. Sheehy and L. Radzihovsky, Annals of Physics {\bf 322}, 1790 (2007) 

\bibitem{iskin} M. Iskin and C. J. Williams, Phys. Rev. A {\bf 77}, 013605, (2008).

\bibitem{blume} D. Blume, Phys. Rev. A {\bf 78}, 013613, (2008).

\end{references}
\end{document}